\documentclass[12pt]{iopart} 

\usepackage{iopams}
\usepackage[dvipdfmx]{graphicx}
\usepackage[dvipdfmx]{color}
\usepackage{dcolumn}
\usepackage{color}
\usepackage{algorithm}
\usepackage{algorithmic}

\newcommand{\red}[1]{\textcolor{red}{#1}}

\begin{document}

\title[NMF for 2D-XAS Images of Lithium Ion Batteries]{Non-Negative Matrix Factorization for 2D-XAS Images of Lithium Ion Batteries}

\author{Hiroki Tanimoto$^1$, Xu Hongkun$^2$, Masaishiro Mizumaki$^3$, Yoshiki Seno$^4$, Jumpei Uchiwada$^5$, Ryo Yamagami$^5$, Hiroyuki Kumazoe$^6$, Kazunori Iwamitsu$^7$, Yuta Kimura$^8$, Koji Amezawa$^8$, Ichiro Akai$^6$, Toru Aonishi$^2$}
\address{$^1$School of Engineering, Tokyo Institute of Technology}
\address{$^2$School of Computing, Tokyo Institute of Technology}
\address{$^3$Japan Synchrotron Radiation Research Institute}
\address{$^4$Saga Prefectural Regional Industry Support Center}
\address{$^5$Graduate School of Science and Technology, Kumamoto University}
\address{$^6$Institute of Industrial Nanomaterials, Kumamoto University}
\address{$^7$Technical Division, Kumamoto University}
\address{$^8$Institute of Multidisciplinary Research for Advanced Materials, Tohoku University}
\ead{aonishi@c.titech.ac.jp}

\vspace{10pt}
\begin{indented}
\item[]October 2021
\end{indented}

\begin{abstract}
Lithium-ion secondary batteries have been used in a wide variety of purposes, such as for powering mobile devices and electric vehicles, but their performance should be improved. One of the factors that limits their performance is the non-uniformity of the chemical reaction in the process of charging and discharging. Many attempts have been made to elucidate the mechanism behind this reaction non-uniformity. In this paper, to detect non-uniformity in various physical properties from Co K-edge two-dimensional X-ray absorption spectroscopy (2D-XAS) images of lithium ion batteries, we propose a method that consists of one-sided orthogonal non-negative matrix factorization in combination with removal of the reference signal. The difference between X-ray absorption spectra acquired at different positions in the battery is very small. However, even in such a situation, our method can decompose the 2D-XAS data into different spatial domains and their corresponding absorption spectra. From the spectral decomposition of the obtained absorption spectra, we confirmed a transition-energy shift of the main peak as evidence for a change in the state of charge and also found spectral changes due to orbital hybridization in the decomposed spectral components.
\end{abstract}

\section{Introduction}
Lithium ion batteries have been used as versatile and effective electric storage systems in a wide range of fields, such as mobile devices, electric vehicles, and energy storage systems for renewable energy. Because high power output and rapid charging are required for progress in such applications, there is a demand for higher performance rates \cite{RN1518,RN1519,RN1520,RN1521}. However, the present lithium ion batteries have poor performance and stability under high rate conditions \cite{RN1519,RN1520,RN1521}. One of the reasons is that they experience spatially non-uniform reactions during charging and discharging. To improve high-rate performance, it is essential to understand the mechanism and the governing factors of the reaction inhomogeneity~\cite{RN1511,RN1512,RN1516}. Especially, two-dimensional (2D) correlation analysis had been applied to 2D X-ray absorption spectroscopy (2D-XAS) and Raman spectra of the Li$_x$CoO system to obtain the detailed information about the electrochemical reaction \cite{RN1639}.

To visualize the reaction inhomogeneity, Nakamura et al. sought to image the lithium content (LC) of Li$_x$CoO$_2$ in a model composite electrode by using 2D-XAS \cite{RN1483}. They two-dimensionally acquired Co-K-edge X-ray absorption spectra of Li$_x$CoO$_2$ composite electrodes. Then, they determined the energy at the highest peak in the acquired spectrum, i.e. peak top energy (PTE) at each of spatial points, and made a two-dimensional map of the LC in the model electrode by using the relation between the PTE and the LC of Li$_x$CoO$_2$ determined from reference materials with well-defined LC values. Through visualization of the reaction inhomogeneity, they showed that the electrochemically active region decreases with increasing current density during charging.

The LC of Li$_x$CoO$_2$ is an important physical factor affecting the uniformity of electrochemical reactions, but other factors such as grain boundaries and structural deformation and the resulting electronic state changes might also contribute to the reaction inhomogeneity. XAS data give not only the LC but also other information on these factors~\cite{RN1501,RN1502,RN1500,NEXAFSSpectroscopy2003,RN1505}. To better understand the mechanism and the governing factors of the reaction inhomogeneity, we need to know the spatial distribution of various physical properties and their spatial correlation. By detecting a spatial domain defined as a set of spatial points with similar spectrum profile from the 2D-XAS data, we can clarify the spatial distribution and the spatial correlation of various physical properties. The purpose of this study was thus to develop a machine-learning method that automatically extracts individual spatial domains with different profiles of absorption spectra from the 2D-XAS data of the model composite electrode measured by Nakamura et al. \cite{RN1483}.

Shiga et al. applied non-negative matrix factorization (NMF) \cite{RN1491} to scanning transmission electron microscopy (STEM) - electron energy-loss spectroscopy (EELS) / energy-dispersive X-ray (EDX) spectral datasets acquired from well-defined reference materials, and they successfully extracted individual domains with different components \cite{RN1495,RN1496}. Furthermore, the NMF method proposed by Shiga et al. has been applied to analyses of STEM-EELS and Raman datasets of lithium-ion-battery electrolyte material \cite{RN1457,RN1474,RN1534}. The profiles of STEM and Raman spectra were clearly distinguishable depending on the dominant components. However, the XAS spectra of the model composite electrode measured by Nakamura et al. are very similar and varied very little from one position to another (see Fig. \ref{fig:gaussfit_full} (C)), because the model composite electrode is composed of the identical compound. As mentioned in Maruyama et al. \cite{RN1497}, a pure NMF, which does not treat the background specially, extracts only a few large signals including the background signal. That is, it is difficult to use NMF to factorize 2D-XAS data of the model composite electrode without removing the common background signal. However, it is difficult to identify the background signal strictly.

In this paper, instead of removing the background signal, the reference signal including the background signal is removed from the 2D-XAS data. In a preprocessing step, we obtain difference spectra by subtracting the X-ray absorption spectrum of a reference material from 2D-XAS data. Li$_{0.5}$CoO$_{2}$ is used as the reference material, which corresponds to a fully charged state~\cite{RN1483}. Subsequently, we apply a one-sided orthogonal NMF (ONMF) to the difference spectra \cite{Park06one-sidednon-negative,RN1492,RN1486}. The one-sided ONMF we propose here imposes column orthogonality on one side's nonnegative factor matrix representing spatial domain structures, but it relaxes the non-negativity restriction on the other side's factor matrix representing the spectra because the difference spectra take signed real values. Furthermore, we estimate the number of spatial domains with different spectral profiles by determining the rank of the factor matrices of the one-sided ONMF by using bi-cross-validation \cite{RN1487}.

The main contribution of this study is the automatic domain extraction method that can identify and distinguish very small differences in X-ray absorption spectra at all positions. To assessed the performance of our method, we used synthetic data that mimic profiles of X-ray absorption spectra. Artificial domains that were used as ground-truth to verify the accuracy of our method were created when synthetic data were generated. We confirmed that the ground-truth domains of the synthetic data were correctly detected and the number of ground-truth domains was correctly estimated. Then, we determined the number of spatial domains with different spectral profiles from real 2D-XAS data of a model composite electrode and extracted spatial domain patterns from the real data. The extracted spatial domains were highly correlated with ones obtained with the k-means method; they were also correlated with patterns obtained by thresholding the PTE map of the X-ray absorption spectra, which reflect the spatial distribution of the LC of Li$_{x}$CoO$_2$. Furthermore, after decomposing the 2D-XAS data into spatial domains and their corresponding spectra, we performed spectral decomposition of the separated X-ray absorption spectra of the respective domains by using Bayesian spectroscopy and found that the spectral features of some of the decomposed spectral components differ in the individual domains \cite{RN1490,RN1494,RN1489}. Hence, this development gives us a means to gain a better understanding of the physical factors governing electrochemical reaction inhomogeneity in the near future.

\section{Materials and Methods}
\subsection{Data and preprocessing}
\begin{figure}[tb]
\begin{center}
\includegraphics[width=10cm]{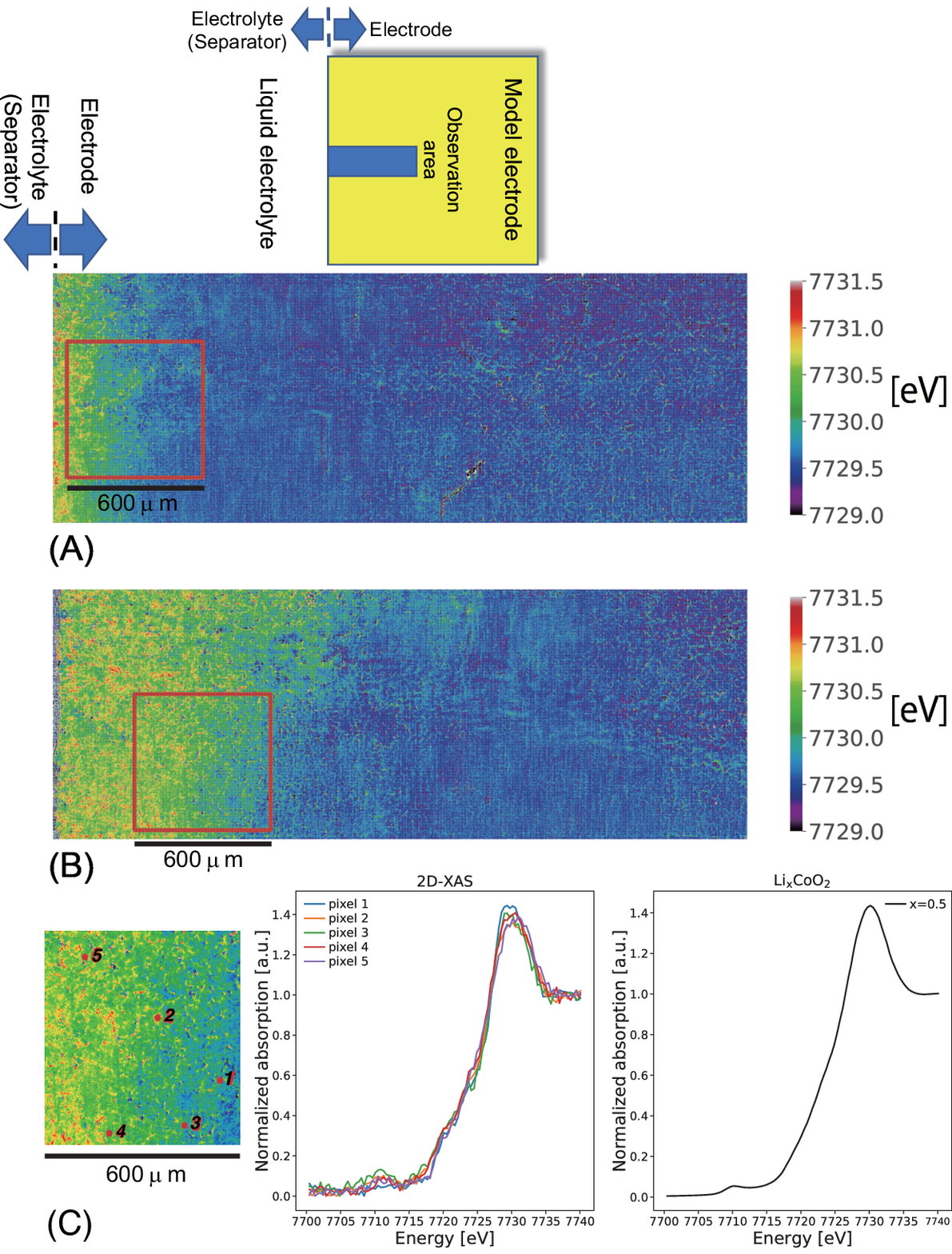}
\caption{2D-XAS data of Co K-edge X-ray absorption spectra of $\mathrm{Li_xCoO_2}$ of the model composite electrode. The two data sets shown in this figure consist of 100 frames (corresponding to an energy range from 7700 to 7740 eV) of $550 \times 1530$ pixels (corresponding to an area of about $1100 \times 3060$ $\mu$m). (A) Two-dimensional map of the PTE obtained from data after charging model electrode to 4.2 V at a current rate of 12 mAcm$^{-2}$. Upper insert: Schematic image of the model electrode and the observation area for 2D-XAS. (B) Two-dimensional map of the PTE obtained from the data after charging the model electrode to 4.2 V at a current rate of 9 mAcm$^{-2}$. The two datasets were $300 \times 300$ pixels (corresponding to an area of about $600 \times 600$ $\mu$m) clipped from the two whole images, as shown by the red squares in (A) and (B). (C) Left panel: The same as the red square area in (B). Middle panel: Spectra at the numbered pixels in the left panel. Right panel: X-ray absorption spectrum of $\mathrm{Li_{0.5}CoO_2}$ reference material. Scale bar = 600 $\mu$m.
}
\label{fig:gaussfit_full}
\end{center}
\end{figure}

The data acquired in Nakamura et al. \cite{RN1483} were provided by the coauthors of this paper. They two-dimensionally observed the change in the Co-K-edge X-ray absorption spectra of Li$_x$CoO$_2$ in the model composite electrode. The XAS measurements at the Co K edge were carried out at the BL01B1 and BL28XU beamlines of SPring-8, Japan. Figures \ref{fig:gaussfit_full} (A) and (B) show two-dimensional maps of the PTE obtained from the data used in this study. An area of about $1100 \times 3060$ $\mu$m was observed with a spatial resolution of approximately 2 $\mu$m in an energy range from 7700 to 7740 eV at energy steps of 0.4 eV. The model electrode was charged to 4.2 V at current rates of 9 mAcm$^{-2}$ (Fig. \ref{fig:gaussfit_full} (B)) and 12 mAcm$^{-2}$ (Fig. \ref{fig:gaussfit_full} (A)). The procedure for fabricating the model electrode and the details of the experiments are given in the papers \cite{RN1499,RN1483}. The two 2D-XAS datasets we analyzed consisted of 100 frames (corresponding to an energy range from 7700 to 7740 eV at energy steps of 0.4 eV) of $300 \times 300$ pixels (corresponding to area of about $600 \times 600$ $\mu$m), which were regions straddling the boundary between the electrochemically active and inactive regions, as shown by the red squares in Figs. \ref{fig:gaussfit_full} (A) and (B).

In the preprocessing, we normalized each Co K-edge X-ray absorption spectrum at each pixel point to be zero at 7700 eV and unity at 7740 eV. Then, we reshaped the 2D array of pixels into a 1D column of pixels for each frame and thereby obtained an $m$ by $n$ spectrum matrix $\boldsymbol{X}_0$. In this study, $m=90000$ and $n=100$. As described in the Introduction, X-ray absorption spectra acquired from the model electrode are very similar and vary only a little from one position to another. To enhance the differences in the absorption spectra, we subtracted a certain reference standard spectrum from the spectrum matrix $\boldsymbol{X}_0$:
\begin{equation}
 \boldsymbol{X} = \boldsymbol{X}_0 - \boldsymbol{1}\boldsymbol{b},
\end{equation}
where $\boldsymbol{1}$ is an $m$-dimensional column vector in which all the elements are $1$, and $\boldsymbol{b}$ is an $n$-dimensional row vector storing the reference standard spectrum. The matrix $\boldsymbol{X}$ obtained by this subtraction is called the difference spectrum matrix. The spectrum matrix $\boldsymbol{X}_0$ is non-negative, whereas the difference spectrum matrix $\boldsymbol{X}$ takes signed values due to the subtraction in Eq. (1). Here, we used the X-ray absorption spectrum of $\mathrm{Li_{0.5}CoO_2}$ reference material (Fig. \ref{fig:gaussfit_full}(C)) as the reference standard spectrum to detect sensitively spectral changes due to charging and discharging, since Li$_{0.5}$CoO$_{2}$ corresponds to the fully charged state~\cite{RN1483}. The reference spectrum in Fig. \ref{fig:gaussfit_full}(C) was separately obtained from the powder sample of Li$_{0.5}$CoO$_2$ through the transmission spectrum measurement with a high intensity beam over a sufficient measurement time.

\subsection{One-sided ONMF for 2D-XAS images}
We assume that the difference spectrum matrix $\boldsymbol{X}$ can be factorized into a real nonnegative matrix $\boldsymbol{W}\in\mathbb{R}_+^{m \times r}$ and a real signed matrix $\boldsymbol{H}\in\mathbb{R}^{n\times r}$ whose rank is equal to $r < \min(m,n)$. Under this assumption, we introduce the following generative model for $\boldsymbol{X}$,
\begin{equation}
\boldsymbol{X} = \boldsymbol{W}\boldsymbol{H}^T + \boldsymbol{\varepsilon},
\end{equation}
where all elements of $\boldsymbol{\varepsilon}\in\mathbb{R}^{m\times n}$ are independent and identically distributed (i.i.d.) Gaussian noise. The $k$-th column vector of $\boldsymbol{W}$, denoted as $\boldsymbol{w}_k\in\mathbb{R}_+^{m\times 1}$, represents the $k$-th spatial domain $(k\in \{1,\cdots,r\})$, and the $k$-th column vector of $\boldsymbol{H}$, denoted as $\boldsymbol{h}_k\in\mathbb{R}^{n \times 1}$, represents the $k$-th difference spectrum corresponding to the $k$-th spatial domain $(k\in \{1,\cdots,r\})$. To represent the difference spectrum taking signed values due to the subtraction in Eq. (1), each element of $\boldsymbol{h}_k$ is defined as a signed real value. Using these column and row vectors, the generative model of Eq. (2) can be rewritten as 
\begin{equation}
\boldsymbol{X} = \sum_{k=1}^{r}\boldsymbol{w}_k\boldsymbol{h}^T_k + \boldsymbol{\varepsilon}.
\end{equation}

Under the i.i.d. Gaussian noise assumption in Eq. (2), $\boldsymbol{W}$ and $\boldsymbol{H}$ can be estimated by minimizing the following objective function,
\begin{equation}
J = ||\boldsymbol{X}-\boldsymbol{W}\boldsymbol{H}^T||_2^2,
\end{equation}
where $||X||_2$ denotes the Frobenius norm.

We seek to minimize the objective function (4) subject to column orthogonality on $\boldsymbol{W}$ by using the hierarchical alternating least squares (HALS) algorithm for the ONMF proposed by Kimura et al. \cite{RN1486}. We summarize the HALS algorithm for one-sided ONMF below.

The HALS algorithm minimizes the objective function defined in Eq. (4) alternately with respect to each of the column vectors $\boldsymbol{w}_k$ and $\boldsymbol{h}_k$ in $\boldsymbol{W}$ and $\boldsymbol{H}$ while keeping the other vectors fixed \cite{RN1484}. Each alternating minimization with respect to $\boldsymbol{w}_k$ and $\boldsymbol{h}_k$ can be equivalently formulated as a minimization of the following objective function,
\begin{eqnarray}
 && J_k = ||\boldsymbol{X}^{(k)} - \boldsymbol{w}_k \boldsymbol{h}_k^T||_2^2, \\
 && \boldsymbol{X}^{(k)} = \boldsymbol{X} - \sum_{k^\prime (\neq k)}\boldsymbol{w}_{k^\prime}\boldsymbol{h}^T_{k^\prime},
\end{eqnarray}
where $\boldsymbol{X}^{(k)}$ denotes the residual between the data $\boldsymbol{X}$ and the factorization result excluding the $k$-th factor $\boldsymbol{w}_k \boldsymbol{h}^T_k$.

Kimura et al. formulated an optimization problem that minimizes $J_k$ subject to the following constraint for imposing column orthogonality on one side's nonnegative factor matrix \cite{RN1486}:
\begin{equation}
 \boldsymbol{W}^{(k)T}\boldsymbol{w}_k = \sum_{k^\prime (\neq k)}\boldsymbol{w}_{k^\prime}^T\boldsymbol{w}_k = 0,
\end{equation}
where $\boldsymbol{W}^{(k)}$ denotes the sum of all column vectors in $\boldsymbol{W}$ excluding the $k$-th column vector $\boldsymbol{w}_{k}$. To solve this optimization problem, they introduced the following Lagrange function with Lagrange multiplier $\lambda_k$,
\begin{equation}
L_k = ||\boldsymbol{X}^{(k)} - \boldsymbol{w}_k\boldsymbol{h}^T_k||_2^2 + \lambda_k(\boldsymbol{W}^{(k)T}\boldsymbol{w}_k).
\end{equation}
Then, they found the stationary points at which the first partial derivatives of $L_k$ with respect to $\boldsymbol{w}_k$ and $\boldsymbol{h}_k$ are zero and obtained an update rule of the HALS algorithm for ONMF. The value of the Lagrange multiplier $\lambda_k$ can be uniquely determined under the column-orthogonality assumption for $W$ \cite{Ding2006}.

The one-sided ONMF we use here imposes column orthogonality on the nonnegative $\boldsymbol{W}$, but relaxes the non-negativity restriction on $\boldsymbol{H}$. To represent the difference spectrum taking signed values, each element of $\boldsymbol{h}_k$ is defined as a real signed value. This is a point of difference from the HALS algorithm proposed by Kimura et al. In our case, the HALS algorithm for the one-sided ONMF is given by the following alternating update equation: 
\begin{eqnarray}
 \boldsymbol{w}_k &\leftarrow& [\boldsymbol{X}^{(k)} \boldsymbol{h}_k
 - \frac{\boldsymbol{W}^{(k)T}\boldsymbol{X}^{(k)}\boldsymbol{h}_k}{\boldsymbol{W}^{(k)T}\boldsymbol{W}^{(k)}}\boldsymbol{W}^{(k)}]_+, \\
 \boldsymbol{h}_k &\leftarrow& {\boldsymbol{X}^{(k)}}^T \boldsymbol{w}_k,
\end{eqnarray}
where $[\cdot{}]_+$ denotes half-wave rectification $[x]_+=\max(x,0)$ resulting in a nonnegative $\boldsymbol{w}_k$. Note that there is no such half-wave rectification on $\boldsymbol{h}_k$, which is different from the HALS algorithm proposed by Kimura et al. \cite{RN1486}. $\lambda_k$ can be set to $\frac{\boldsymbol{W}^{(k)T}\boldsymbol{X}^{(k)}\boldsymbol{h}_k}{\boldsymbol{W}^{(k)T}\boldsymbol{W}^{(k)}}$ as in Kimura et al. \cite{RN1486}. Algorithm \ref{alg1} is the HALS algorithm for the one-sided ONMF.

In the preprocessing, we obtain difference spectra by subtracting the reference standard spectrum from 2D-XAS data. After the preprocessing, we apply the one-sided ONMF to the difference spectra matrix $\boldsymbol{X}$ as described above. Thus, $\boldsymbol{H}$ represents background-subtracted spectra of the extracted domains. To obtain original X-ray absorption spectra of the extracted domains, we calculate $\boldsymbol{W}^T \boldsymbol{X}_0$. Because the one-sided ONMF imposes column orthogonality on one side's nonnegative matrix $\boldsymbol{W}$ representing individual domains, $\boldsymbol{W}^T \boldsymbol{W} \simeq I$ is satisfied, and thus calculating $\boldsymbol{W}^T \boldsymbol{X}_0$ can approximately give the X-ray absorption spectra of the individual domains. This calculation corresponds to a weighted average of the X-ray absorption spectra for each domain.

\begin{algorithm}
\caption{HALS algorithm for one-sided ONMF}
\label{alg1}
\begin{algorithmic}[1]
\STATE Initialize $\boldsymbol{W}$ and $\boldsymbol{H}$ randomly.
\STATE $\boldsymbol{U} = \boldsymbol{W} \boldsymbol{1}_r$
\WHILE{Convergence criterion is not satisfied}
\STATE $\boldsymbol{A} = \boldsymbol{X} \boldsymbol{H}$
\STATE $\boldsymbol{B} = \boldsymbol{H}^T \boldsymbol{H}$
\FOR{k=1 to r}
\STATE $\boldsymbol{W}^{(k)} = \boldsymbol{U} - \boldsymbol{w}_k$
\STATE $u = \boldsymbol{A}_{: k} - \boldsymbol{W}\boldsymbol{B}_{: k}+\boldsymbol{B}_{k k}\boldsymbol{w}_k$
\STATE $\boldsymbol{w}_k = [\boldsymbol{u}
 - \frac{\boldsymbol{W}^{(k)T}\boldsymbol{u}}{\boldsymbol{W}^{(k)T}\boldsymbol{W}^{(k)}}\boldsymbol{W}^{(k)}]_+$
\STATE $\boldsymbol{w}_k = \boldsymbol{w}_k/||\boldsymbol{w}_k ||^2$
\STATE $\boldsymbol{U} = \boldsymbol{W}^{(k)} + \boldsymbol{w}_k$
\ENDFOR
\STATE $\boldsymbol{C} = \boldsymbol{W}^T \boldsymbol{X}$
\STATE $\boldsymbol{D} = \boldsymbol{W}^T \boldsymbol{W}$
\FOR{k=1 to r}
\STATE $\boldsymbol{h}_{k} = \boldsymbol{C}_{: k} - \boldsymbol{H}\boldsymbol{D}_{: k} + \boldsymbol{D}_{kk} \boldsymbol{h}_k $
\ENDFOR
\ENDWHILE
\end{algorithmic}
\end{algorithm}

\subsection{Bi-cross validation for model selection}
We seek to estimate the number of spatial domains with different spectral profiles by applying model selection to the one-sided ONMF. Here, we can use bi-cross-validation \cite{RN1487} to determine the rank of the two factor matrices from the data $X$. Bi-cross-validation is applicable to a wide class of matrix factorization methods \cite{RN1487}. 

The data $\boldsymbol{X}\in\mathbb{R}^{m\times n}$ are separated into four blocks consisting of $\boldsymbol{X}_A$, $\boldsymbol{X}_B$, $\boldsymbol{X}_C$ and $\boldsymbol{X}_D$, as follows:
\begin{equation}
 \boldsymbol{X} = \left[
 \begin{array}{cc}
 \boldsymbol{X}_A & \boldsymbol{X}_B\\
 \boldsymbol{X}_C & \boldsymbol{X}_D
 \end{array}\right].
 \label{eq:BCV}
\end{equation}
Let $\boldsymbol{X}_D$ be factorized into $\boldsymbol{W}_D$ and $\boldsymbol{H}_D$ by applying the HALS algorithm described above. The estimated value of the denoised $\boldsymbol{X}_D$ becomes $\hat{\boldsymbol{X}}_D=\boldsymbol{W}_D\boldsymbol{H}_D^T$. As reported in \cite{RN1487}, the generalization error can be measured from a residual defined by $\hat{\boldsymbol{\varepsilon}}_A = \boldsymbol{X}_A - \boldsymbol{X}_B \hat{\boldsymbol{X}_D}^{+} \boldsymbol{X}_C$, where $\hat{\boldsymbol{X}_D}^{+}$ is the pseudo-inverse of $\hat{\boldsymbol{X}_D}$.

The detailed procedure for calculating the cross-validation error is as follows. Every 20th row and column vectors are selected from $\boldsymbol{X}$. The partial blocks $\boldsymbol{X}_A$, $\boldsymbol{X}_B$, $\boldsymbol{X}_C$ and $\boldsymbol{X}_D$ are organized with these picked vectors and the remaining row and column vectors. Then, $\boldsymbol{X}_D$ is factorized into $\boldsymbol{W}_D$ and $\boldsymbol{H}_D$ by the HALS described above, and the generalization error $||\hat{\boldsymbol{\varepsilon}}_A||_2^2$ is calculated. This process is repeated 20 times until all the row and column vectors have been picked, and the average of $||\hat{\boldsymbol{\varepsilon}}_A||_2^2$ is obtained. Hereafter, the average of $||\hat{\boldsymbol{\varepsilon}}_A||_2^2$ is called the bi-cross-validation error. Then, the rank of the two factor matrices was determined by searching for the minimum point of the bi-cross-validation error.

\subsection{Synthetic data}
\begin{figure}[tb]
\begin{center}
\includegraphics[width=12cm]{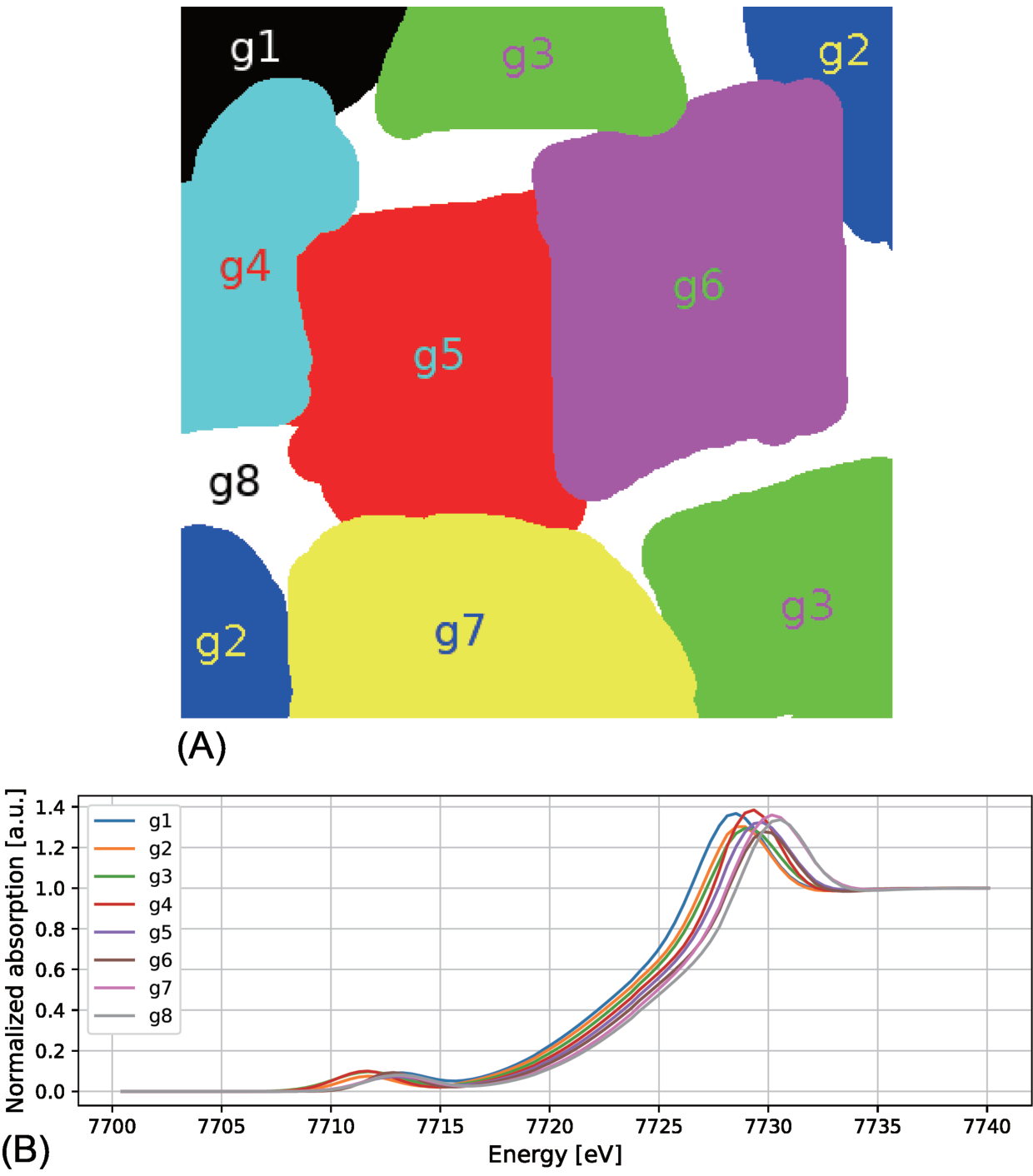}
\caption{Synthetic data. (A) Eight separate domains g1--g8 drawn by hand. (B) Eight different spectra g1--g8 mimicking profiles of X-ray absorption spectra.}
\label{fig:demo_image}
\end{center}
\end{figure}

We generated synthetic data as ground truth on which we could objectively assess the performance of our method.

We manually created eight 400 by 400 binary arrays representing eight separate domains (Fig. \ref{fig:demo_image}(A)) and stored them as a 160000 by 8 binary matrix $\boldsymbol{W}_{g}$ by reshaping the 2D binary array of each of these domains into a 1D binary array. We synthesized eight different spectra on a 100 by 8 matrix $\boldsymbol{H}_{g}$, in which each consisted of a randomly shifted single error function and two Gaussians to mimic profiles of X-ray absorption spectra (Fig. \ref{fig:demo_image}(B)). Then, we obtained a 160000 by 100 synthetic spectrum matrix $\boldsymbol{X}_0$ by multiplying $\boldsymbol{W}_{g}$ and $\boldsymbol{H}_{g}^T$ and adding Gaussian noise with SN ratio of 5.62dB. We also created a spectrum mimicking the X-ray absorption spectrum of the $\mathrm{Li_{0.5}CoO_2}$ reference material in the same manner. 

\subsection{k-means method}
To confirm the consistency of the results obtained with the one-sided ONMF and other methods, we used the k-means method to analyze the 2D-XAS data and compared its results with those of our method. The k-means method is a standard clustering method that classifies data into $k$ clusters. To extract individual spatial domains with different spectral profiles, it was formulated as an optimization problem \cite{RN1488}:
\begin{equation}
 \min_{\{\boldsymbol{h}_j\}, \{w_{ij}\}}\sum_{i=1}^m \sum_{j=1}^r w_{ij}||\boldsymbol{x}_i - \boldsymbol{h}_j||^2_2,
\end{equation}
where $\boldsymbol{x}_i$ is the X-ray absorption spectrum at the $i$-th pixel, and $\boldsymbol{h}_j$ denotes the mean absorption spectrum of the $j$-th domain. $w_{ij}$ is a one-hot representation that takes either $1$ or $0$ depending on whether or not $\boldsymbol{x}_i$ belongs to the $j$-th domain. The vector denoted by $\boldsymbol{w}_i^{k-means}$ represents the spatial structure of the $i$-th domain .

To quantitatively compare results obtained with the one-sided ONMF and those with the k-means method, we calculated the cross correlation of all combinations of $\boldsymbol{w}_{i'}$ of the one-sided ONMF and $\boldsymbol{w}_i^{k-means}$,
\begin{equation}
 \mathrm{Corr}_{i',i} =
 \frac{\boldsymbol{w}_{i'}^T \boldsymbol{w}_{i}^{k-means}}{\sqrt{||\boldsymbol{w}_{i'}||_2||\boldsymbol{w}_{i}^{k-means}||_2}}. \label{eq:corr}
\end{equation}

\subsection{Peak top energy}
To visualize the reaction inhomogeneity, Nakamura et al. obtained two-dimensional maps of LC of Li$_x$CoO$_2$ in the model composite electrode from the PTE of the X-ray absorption spectrum at each pixel \cite{RN1483}. Here, we confirmed the consistency of the LC map and the domain patterns obtained by the one-sided ONMF. Because the LC has a one-to-one correlation with the PTE between 7728.5$[\mathrm{eV}]$ $<$ PTE $<$ 7730.2$[\mathrm{eV}]$, instead of comparing the LC map and the domain patterns, we compared the PTE map with the domain patterns.

First, we determined the PTE by fitting a Gaussian distribution to the X-ray absorption spectrum at each pixel and obtained two-dimensional maps of the PTE. Next, we classified pixels on the obtained map into three sets, denoted by D1, D2 and D3, which have PTEs in the ranges of $(E_2,\infty)$, $(E_2,E_1)$ and $(0,E_1)$, respectively. $E_1$ and $E_2$ are thresholds for classifying all of the pixels into D1, D2, or D3. 

To quantitatively compare results obtained with the one-sided ONMF and those with the PTE, we calculated the cross correlation of $\boldsymbol{w}_{i}$ of the one-sided ONMF and pixels of Dk using Eq. (\ref{eq:corr}). In this comparison, the threshold $E_1$ was determined by maximizing the cross correlation between the pixels of D3 and the most similar $\boldsymbol{w}_{i}$ to the pixels of D3, while the threshold $E_2$ was determined by maximizing the cross correlation between the pixels of D1 and the most similar $\boldsymbol{w}_{i}$ to the pixels of D1.

Furthermore, to confirm the consistency between results obtained with the one-sided ONMF and with respect to the PTE, we compared the PTEs of the three domains of the one-sided ONMF that were most highly correlated with D1, D2 and D3 with the values of the thresholds $E_1$ and $E_2$.

\subsection{Bayesian spectroscopy for X-ray absorption spectra}
In order to reveal the physical property changes in the respective domains, we performed Bayesian spectroscopy~\cite{RN1490,RN1494,RN1489} on the extracted X-ray absorption spectra (XAS). Bayesian spectroscopy can decompose XAS into an absorption edge step structure, a white line (WL) characterized with the PTE, pre-edge structures with weak absorption intensities that appear on the low-energy side of the step structure, as well as and other peak structures \cite{RN1489}. Furthermore, by using the Bayes free energy as an information criterion~\cite{RN1490}, we can estimate the number of spectral components for the pre-edge and other peak structures through this data-driven science approach.

To decompose the XAS, Eq.~(\ref{iakaiEq01}) was used as a model that is a function of photon energy $E$.

\begin{equation}
  \mathcal{F}_{\red{K}}(E;\boldsymbol{\theta})
 := y_{0}+
 \mathcal{S}(E;\boldsymbol{\theta}_\mathcal{S})
 +
 \sum_{k=1}^{K}
 \mathcal{G}_k(E;\boldsymbol{\theta}_k)
 ,
 \label{iakaiEq01}
\end{equation}
where bold symbols denote the spectral parameters for the respective spectral components and $y_{0}$ is an energy-independent baseline to correct for zero. The spectral function $\mathcal{S}(E;\boldsymbol{\theta}_\mathcal{S})$ consists of a gentle step and a WL peak, as defined in Eq.~(\ref{iakaiEq02}).

\begin{eqnarray}
 \mathcal{S}(E;\boldsymbol{\theta}_\mathcal{S})
 & := &
 H
 \left[
 \frac{1}{2}
 +
 \frac{1}{\pi}
 \arctan
 \left(
 \frac{
 E - E_0
 }{
 {\it\Gamma}/2
 }
 \right)
 \right]
 \nonumber
 \\
 & &
 \hspace*{1em}
 +
 \mathcal{V}(E;A,E_\textrm{WL},\omega,\eta)
 ,
 \label{iakaiEq02}
\end{eqnarray}
where the first term is a step structure (an arctangent function) \cite{NEXAFSSpectroscopy2003} characterized by an intensity $H$, the transition energy of the X-ray absorption edge $E_0$, and a broadening factor ${\it\Gamma}$. The WL giving the PTE is represented by the second term in Eq.~(\ref{iakaiEq02}), the pseudo-Voigt function~\cite{JApplCryst1986V19p63-64} parameterized by an integrated intensity $A$, the transition energy $E_\textrm{WL}$, the spectral width $\omega$, and the mixing ratio $\eta$ between Lorentz and Gaussian shapes. Since the WL appears near the absorption edge energy $E_{0}$, a normal distribution with a standard deviation of 4.5~eV is introduced for a prior probability of the energy difference between $E_{0}$ and $E_\textrm{WL}$.
The pre-edge and other peak structures are simplified and represented as a sum of Gaussian shapes $\mathcal{G}_k(E;\boldsymbol{\theta}_k)$ ($k=1,\cdots,K$), as in the second term of Eq.~(\ref{iakaiEq01}).

In Bayesian spectroscopy~\cite{RN1490,RN1494}, we apply the Bayes' theorem~\cite{PhilTrans1763V53p370} to the
spectral decomposition analysis of the spectral data
$\boldsymbol{D}:=\{(E_1,y_1),\cdots,(E_i,y_i),\cdots,(E_N,y_N)\}$. The model
$\mathcal{F}_{K}(E_i;\boldsymbol{\theta})$ in Eq.~(\ref{iakaiEq01}) is specified by a parameter set
$\boldsymbol{\theta}$, and according to the causality, the cause $\boldsymbol{\theta}$ characterizes the
resultant $\boldsymbol{D}$. The joint probability of $\boldsymbol{\theta}$ and $\boldsymbol{D}$,
$P(\boldsymbol{\theta},\boldsymbol{D})$, can be expanded as
\begin{math}
  P(\boldsymbol{\theta},\boldsymbol{D})
  =
  P(\boldsymbol{D}|\boldsymbol{\theta})P(\boldsymbol{\theta})
  =
  P(\boldsymbol{\theta}|\boldsymbol{D})P(\boldsymbol{D})
  ,
\end{math}
while Bayesian inference evaluates the posterior (conditional) probability distribution of the cause
$\boldsymbol{\theta}$ in Eq.~(\ref{Eq:Bayes01}) under the condition $\boldsymbol{D}$ given.  
\begin{equation}
  P(\boldsymbol{\theta}|\boldsymbol{D})
  =
  \frac{
  P(\boldsymbol{D}|\boldsymbol{\theta})P(\boldsymbol{\theta})
  }{
  P(\boldsymbol{D})
  }
  ,
  \label{Eq:Bayes01}
\end{equation}
When the noise superimposed on the data ($\{\cdots,y_i,\cdots\}$)
is normally distributed, the likelihood term, $P(\boldsymbol{D}|\boldsymbol{\theta})$, is written in the
following equation:
\begin{displaymath}
  P(\boldsymbol{D}|\boldsymbol{\theta})
  =
  \left(
  \frac{b}{2\pi}
  \right)^{N/2}
  \exp
  \left[
  -bN\mathcal{E}_{K}(\boldsymbol{\theta})
  \right]
  ,
\end{displaymath}
where $b$ is a quasi-inverse temperature defined as $b:=\sigma_\textrm{noise}^{-2}$ with the standard deviation
of the imposed noise on the data, and $\mathcal{E}(\boldsymbol{\theta})$ is defined as follows: 
\begin{displaymath}
  \mathcal{E}_{K}(\boldsymbol{\theta})
  :=
  \frac{1}{2N}
  \sum_{i=1}^{N}
  \left[
  y_i
  -
  \mathcal{F}_{K}(E_i;\boldsymbol{\theta})
  \right]^2
\end{displaymath}
and is an error function of the model $\mathcal{F}_{K}(E_i;\boldsymbol{\theta})$ specified by the parameter set
$\boldsymbol{\theta}$.  The denominator in Eq.~(\ref{Eq:Bayes01}) is a Bayesian partition
function~\cite{RN1490}, and is obtained by marginalization of the numerator terms of Eq.~(\ref{Eq:Bayes01}) in
parameter space $\boldsymbol{\theta}$.
Model selection in Bayesian spectroscopy is performed using the Bayesian free energy as the information
criterion~\cite{RN1490}. In the model of Eq.~(\ref{iakaiEq01}), the number of components $K$ of the Gaussian
line-shape is estimated by model selection, and the Bayesian partition function becomes $Z(K,b)$, a function of
the number of components $K$ and the quasi-inverse temperature $b$.  The Bayesian free energy is defined as
$F(K,b):=-\ln{}Z(K,b)$~\cite{RN1490}, and the model selection~\cite{RN1490} and the noise
inference~\cite{JPSJ2017V86p024001} can be performed as follows:
\newcommand{\argmin}{\mathop{\rm arg~min}\limits}
\newcommand{\argmax}{\mathop{\rm arg~max}\limits}
\begin{displaymath}
  (\hat{K},\hat{b})
  =
  \argmin_{K,b}
  F(K,b).
\end{displaymath}
the maximum a posteriori (MAP) estimation of the parameter set $\boldsymbol{\theta}$ is performed by posterior probability maximization as following:
\begin{displaymath}
  \hat{\boldsymbol{\theta}}
  =
  \argmax_{\boldsymbol{\theta}}
  \left\{
  \exp
  \left[
  -\hat{b}N\mathcal{E}_{\hat{K}}(\boldsymbol{\theta})
  \right]
  P(\boldsymbol{\theta})
  \right\}
  .
\end{displaymath}

\section{Results}
\subsection{Performance evaluation using synthetic data}
\begin{figure}[tb]
\begin{center}
\includegraphics[width=12cm]{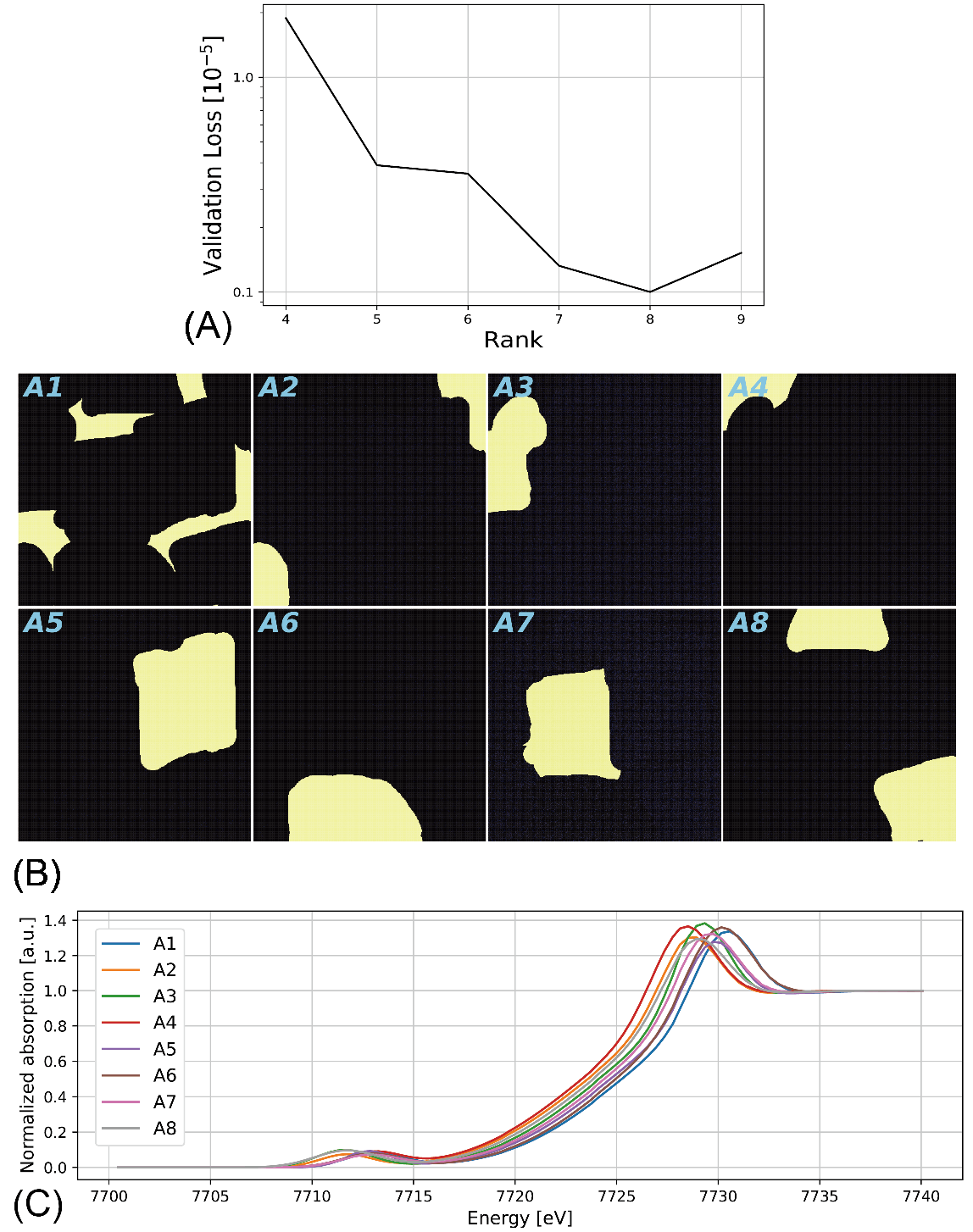}
\caption{Performance evaluation of one-sided ONMF on synthetic data. (A) Bi-cross-validation error as a function of rank $r$ for $4 \leq r \leq 9$. (B) Eight spatial domains A1--A8 factorized by one-sided ONMF. (C) Absorption spectra corresponding to domains A1--A8.}
\label{fig:demo_NMF}
\end{center}
\end{figure}

We assessed the performance of our method on synthetic data with eight separate domains.

First, the rank of the factor matrices $r$ was determined by bi-cross-validation. Figure \ref{fig:demo_NMF}(A) shows the bi-cross-validation error as a function of rank $r$ for $4 \leq r \leq 9$. The bi-cross-validation error was minimized at $r=8$; thus, the estimated rank was confirmed to be identical to the number of ground-truth domains of synthetic data.

When the rank of the factor matrices was set to eight $(r=8)$, the one-sided ONMF successfully extracted eight spatial domains with different spectral profiles from the synthetic data (Figs. \ref{fig:demo_NMF}(B) and (C)). A1--A8 in Fig. \ref{fig:demo_NMF}(B) show spatial domain patterns obtained by reshaping the column vectors of the factorized matrix $\boldsymbol{W}$ into 2D arrays, and A1--A8 in Fig. \ref{fig:demo_NMF}(C) indicate the absorption spectra corresponding to domains A1--A8. A comparison of A1--A8 and g1--g8 in Fig. \ref{fig:demo_image} confirmed that the spatial domains obtained by our method were identical with the ground-truth domains g1--g8 in the synthetic data.

\subsection{Analyses of real 2D-XAS data}
\begin{figure}[tb]
\begin{center}
\includegraphics[width=12cm]{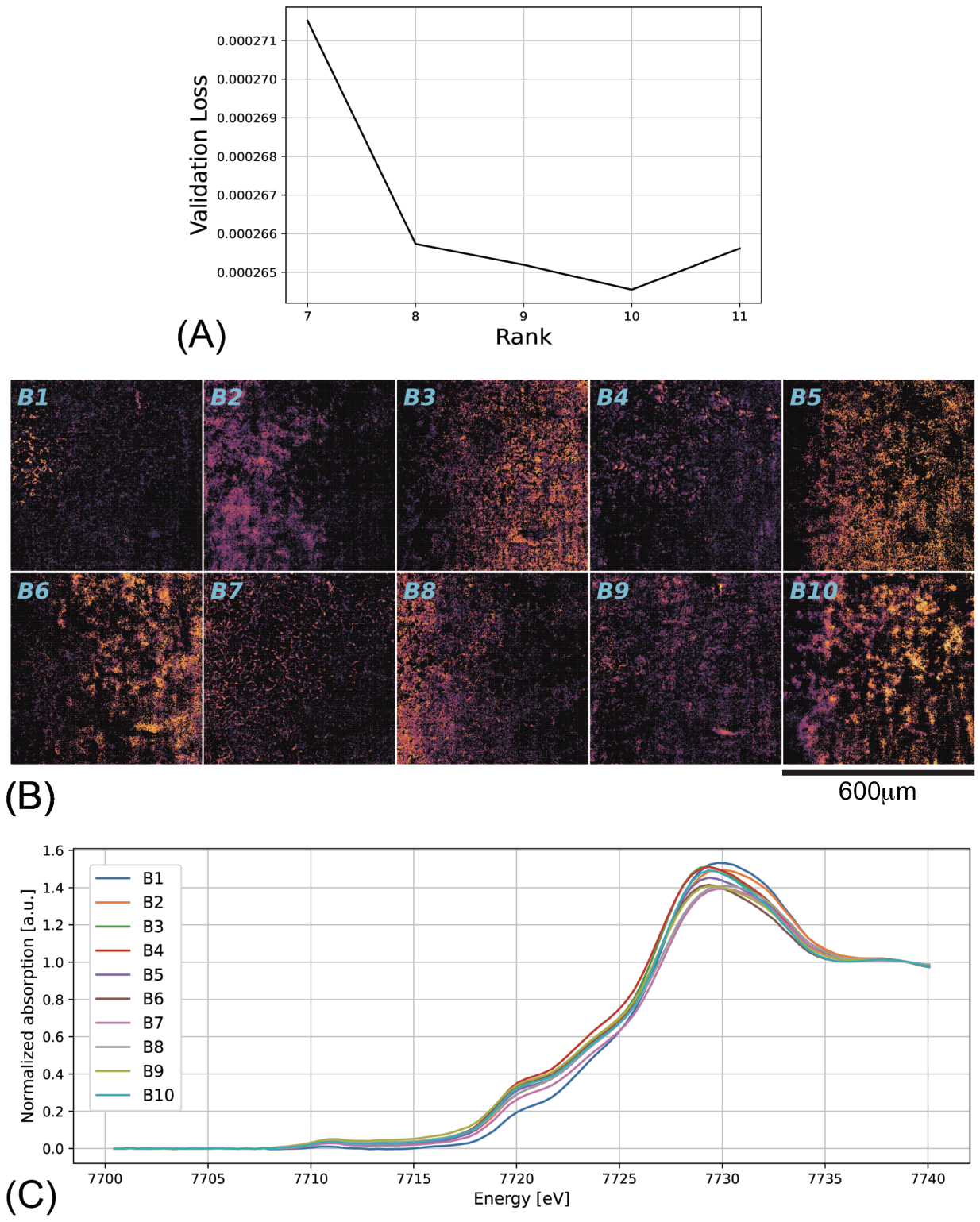}
\caption{Extraction of spatial domains and spectra from real 2D-XAS data of model electrode charged to 4.2 V at current rate of 12 mAcm$^{-2}$ in Fig. \ref{fig:gaussfit_full}(A) by using one-sided ONMF in combination with removal of the reference signal. (A) Bi-cross-validation error as a function of rank $r$ for $7 \leq r \leq 11$. (B) Ten spatial domains B1--B10 factorized by the one-sided ONMF. Scale bar = 600 $\mu$m. (C) Absorption spectra corresponding to domains B1--B10.}
\label{fig:XAFS_NMF1}
\end{center}
\end{figure}

\begin{figure}[tb]
\begin{center}
\includegraphics[width=12cm]{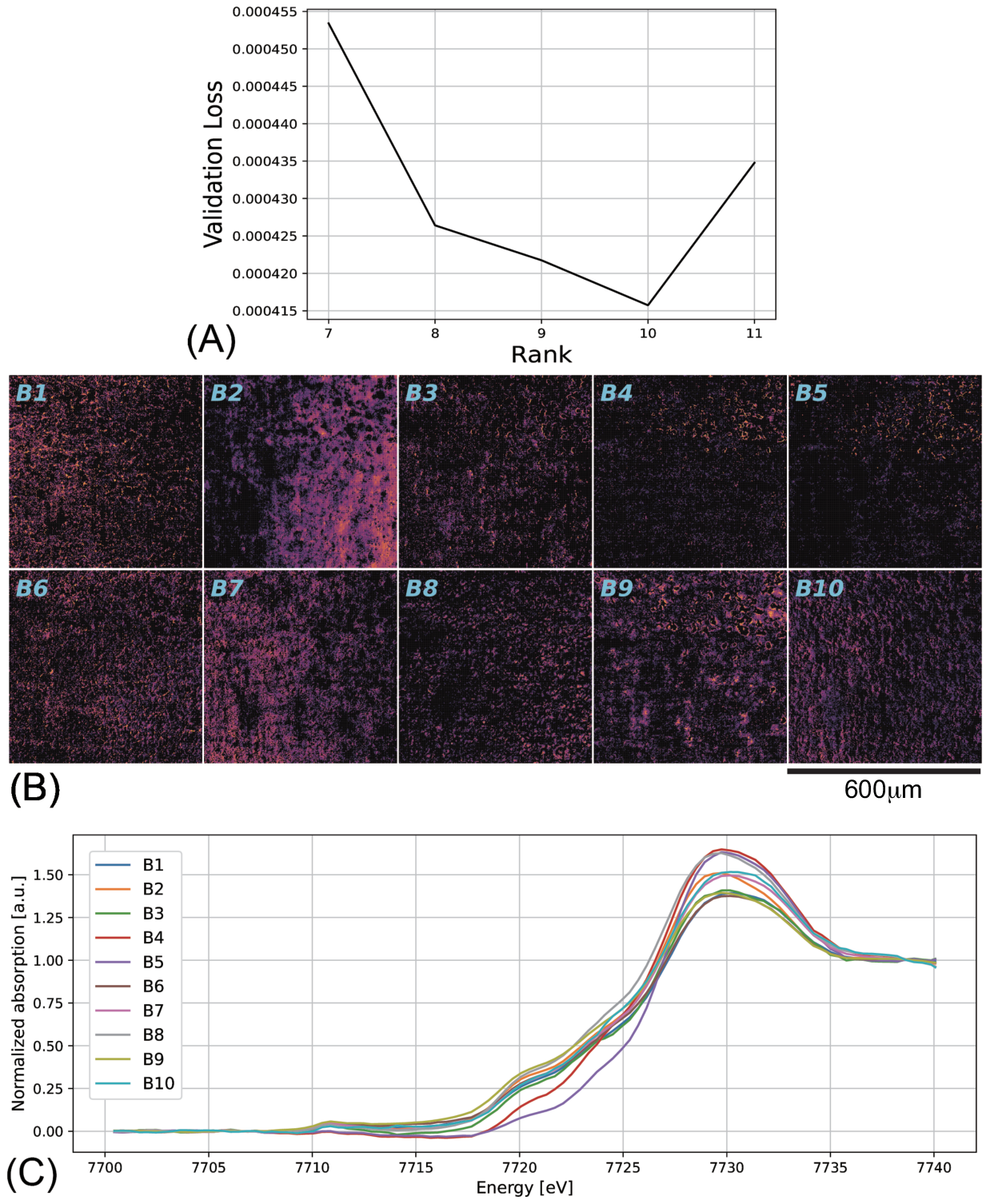}
\caption{Extraction of spatial domains and spectra from real 2D-XAS data of model electrode charged to 4.2 V at current rate of 9 mAcm$^{-2}$ in Fig. \ref{fig:gaussfit_full}(B) by using one-sided ONMF in combination with removal of the reference signal. (A) Bi-cross-validation error as a function of rank $r$ for $7 \leq r \leq 11$. (B) Ten spatial domains B1--B10 factorized by the one-sided ONMF. Scale bar = 600 $\mu$m. (C) Absorption spectra corresponding to domains B1--B10.} 
\label{fig:XAFS_NMF2}
\end{center}
\end{figure}

By using the one-sided ONMF in combination with removal of the reference signal, we sought to extract spatial domains with different spectral profiles from the real 2D-XAS data of the model composite electrode. Here, we used two data sets for the model electrode charged to 4.2 V at current rates of 9 mA cm$^{-2}$ (Fig. \ref{fig:gaussfit_full} (B)) and 12 mA cm$^{-2}$ (Fig. \ref{fig:gaussfit_full} (A)) measured by Nakamura et al.\cite{RN1483}. To reduce the calculation time, we clipped data from the whole images, as indicated by the squares in Figs. \ref{fig:gaussfit_full} (A) and (B).

First, the rank of the factor matrices $r$ was determined by bi-cross-validation. Figures \ref{fig:XAFS_NMF1}(A) and \ref{fig:XAFS_NMF2}(A) show the bi-cross-validation error as a function of rank $r$ for $7 \leq r \leq 11$. The bi-cross-validation errors were minimized at $r=10$ on both data.

Next, we examined how our method decomposed the 2D-XAS data into spatial domains and their corresponding absorption spectra when the rank of the factor matrices was set to ten $(r=10)$ (Figs. \ref{fig:XAFS_NMF1} and \ref{fig:XAFS_NMF2}). B1--B10 in Figs. \ref{fig:XAFS_NMF1}(B) and \ref{fig:XAFS_NMF2}(B) show spatial domain patterns obtained by reshaping the column vectors of the factorized matrix $\boldsymbol{W}$, and B1--B10 in Figs. \ref{fig:XAFS_NMF1}(C) and \ref{fig:XAFS_NMF2}(C) indicate absorption spectra corresponding to the ten domains B1--B10 in Figs. \ref{fig:XAFS_NMF1}(B) and \ref{fig:XAFS_NMF2}(B).

\subsection{Comparison with k-means method}
\begin{figure}[tb]
\begin{center}
\includegraphics[width=12cm]{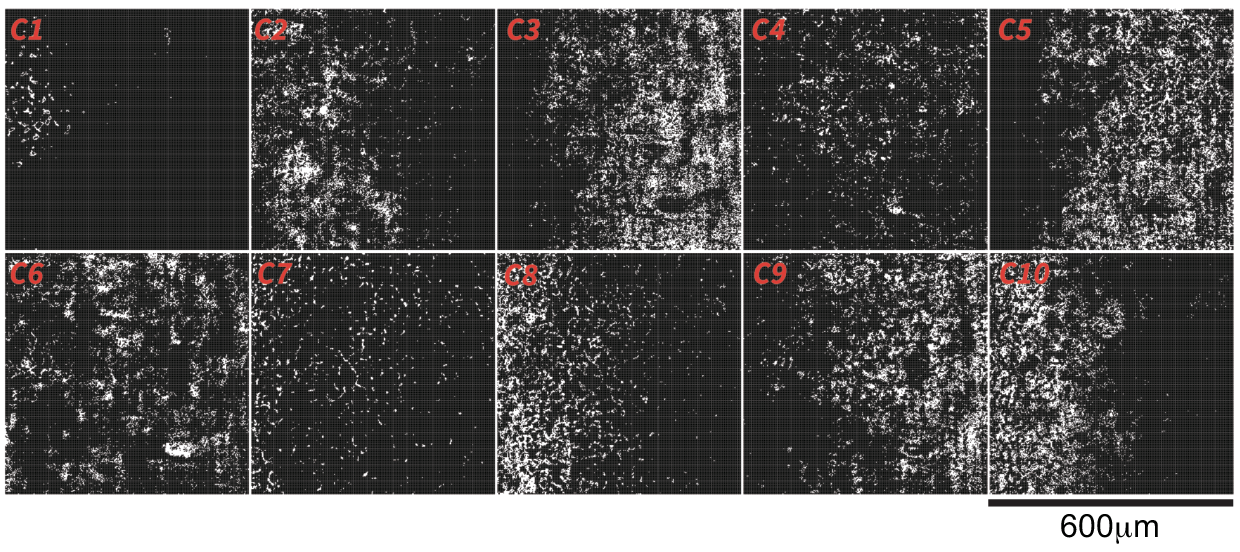}
\caption{Extraction of ten spatial domains C1--C10 using k-means method from real 2D-XAS data of model electrode charged to 4.2 V at current rate of 12 mAcm$^{-2}$ in Fig. \ref{fig:gaussfit_full}(A). Scale bar = 600 $\mu$m.}
\label{fig:XAFS_KMeans_W}
\end{center}
\end{figure}
\begin{table}[tb]
\begin{center}
\caption{Cross correlation matrix of domain patterns extracted with one-sided ONMF and k-means method. Each element is a cross correlation between each of B1--B10 in Fig. \ref{fig:XAFS_NMF1}(B) and each of C1--C10 in Fig. \ref{fig:XAFS_KMeans_W}.}
\label{tab:correlation_NMF_KMeans}
 
\begin{tabular}{|l|rrrrrrrrrr|}
 \hline
 {} & C1 & C2 & C3 & C4 & C5 & C6 & C7 & C8 & C9 & C10 \\
 \hline
 B1 & {\bf 0.80} & 0.11 & 0.03 & 0.08 & 0.05 & 0.06 & 0.02 & 0.10 & 0.00 & 0.01 \\
 B2 & 0.03 & {\bf 0.61} & 0.07 & 0.18 & 0.03 & 0.03 & 0.02 & 0.22 & 0.01 & 0.28 \\
 B3 & 0.01 & 0.09 & {\bf 0.59} & 0.33 & 0.05 & 0.02 & 0.01 & 0.03 & 0.24 & 0.04 \\
 B4 & 0.07 & 0.05 & 0.07 & {\bf 0.53} & 0.03 & 0.03 & 0.01 & 0.00 & 0.25 & 0.07 \\
 B5 & 0.00 & 0.12 & 0.12 & 0.02 & {\bf 0.56} & 0.06 & 0.01 & 0.12 & 0.14 & 0.14 \\
 B6 & 0.02 & 0.03 & 0.07 & 0.01 & 0.16 & {\bf 0.55} & 0.08 & 0.00 & 0.31 & 0.01 \\
 B7 & 0.08 & 0.04 & 0.03 & 0.01 & 0.09 & 0.07 & {\bf 0.54} & 0.32 & 0.01 & 0.03 \\
 B8 & 0.01 & 0.04 & 0.05 & 0.01 & 0.07 & 0.13 & 0.28 & {\bf 0.48} & 0.01 & 0.32 \\
 B9 & 0.06 & 0.02 & 0.08 & 0.02 & 0.04 & {\bf 0.37} & 0.10 & 0.21 & 0.34 & 0.13 \\
 B10& 0.02 & 0.12 & 0.31 & 0.16 & {\bf 0.32} & 0.07 & 0.05 & 0.09 & 0.15 & 0.08 \\
 \hline
 \end{tabular}
 \end{center}
\end{table}

To confirm the consistency of the results obtained with the one-sided ONMF and the other methods, the k-means method was used to classify real 2D-XAS data into several clusters. Here, we used the data from the model electrode charged to 4.2 V at a current rate of 12 mAcm$^{-2}$ (Fig. \ref{fig:gaussfit_full} (A)). In accordance with the results of the model selection for the one-sided ONMF, we separated the 2D-XAS data into ten clusters by using the k-means method (see Fig. \ref{fig:XAFS_KMeans_W}).

To quantitatively compare results obtained with the one-sided ONMF and with the k-means method, we calculated the cross correlation of the domain patterns extracted by both methods. Table \ref{tab:correlation_NMF_KMeans} shows the cross correlation matrix between B1--B10 in Fig. \ref{fig:XAFS_NMF1}(B) and C1--C10 in Fig. \ref{fig:XAFS_KMeans_W}. Domain patterns B1--B8 obtained with the one-sided ONMF were highly correlated with C1--C8 obtained with the k-means method.

\subsection{Comparison with spatial domain patterns with respect to PTE}\label{Sec03-04}
\begin{figure}[tb]
\begin{center}
\includegraphics[width=9cm]{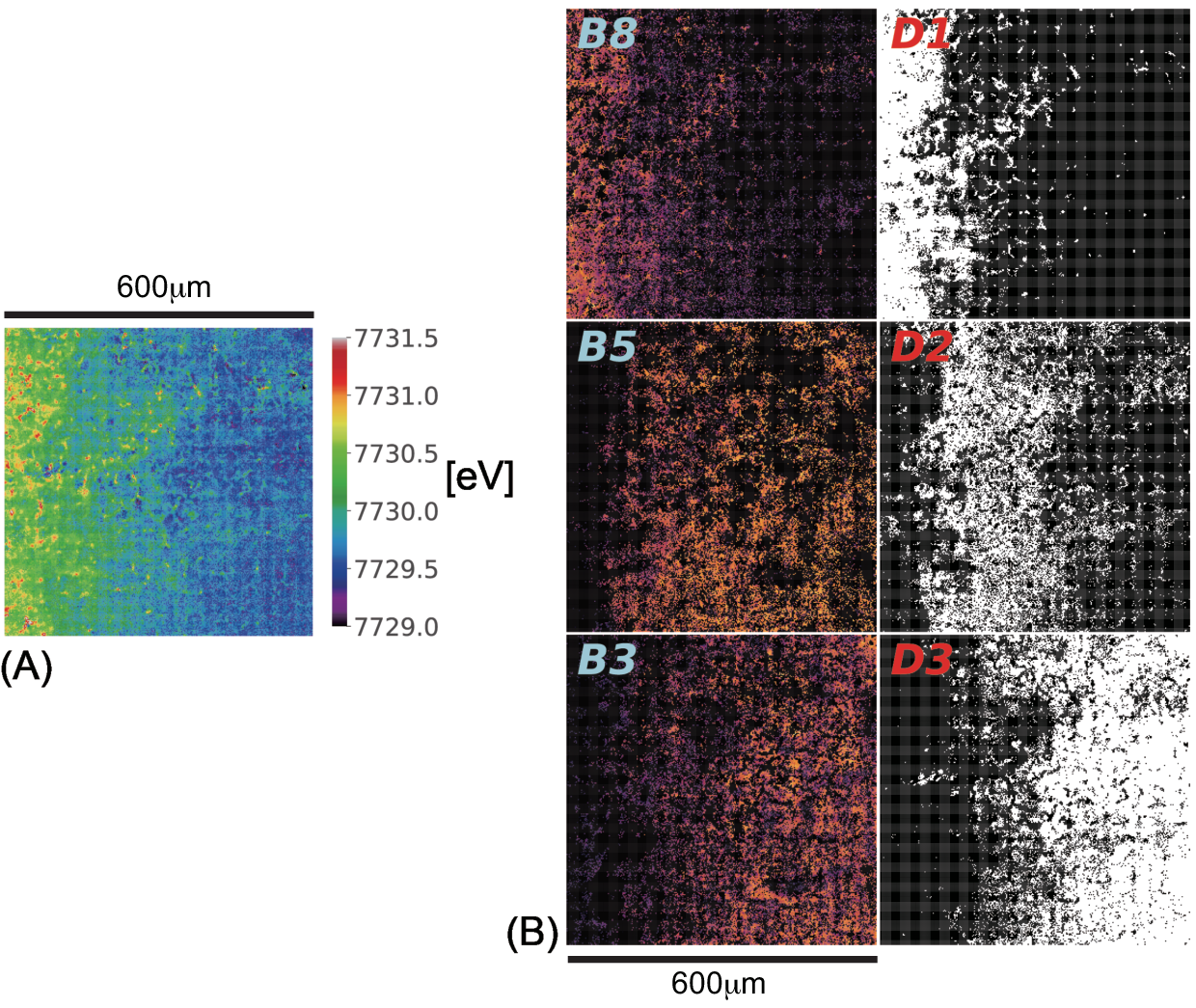}
\caption{Comparison of spatial domain patterns obtained by one-sided ONMF and by thresholding the PTE map. (A) Two-dimensional map of the PTE for model electrode charged to 4.2 V at current rate of 12 mAcm$^{-2}$ in Fig. \ref{fig:gaussfit_full}(A). (B) D1, D2, D3: Three sets of pixels classified with respect to the PTE by thresholds $E_1$ and $E_2$. $E_1 = 7730.18[\mathrm{eV}]$ and $E_2 = 7729.79[\mathrm{eV}]$. B8, B5, B3: Three domain patterns of one-sided ONMF most highly correlated with D1, D2 and D3. Scale bar = 600 $\mu$m.}
\label{fig:gaussfit}
\end{center}
\end{figure}
\begin{table}[tb]
\begin{center}
\caption{Cross correlation matrix of domain patterns obtained by one-sided ONMF and by thresholding the PTE map. Each element is a cross correlation between each of B1--B10 in Fig. \ref{fig:XAFS_NMF1} (B) and each of D1--D3 in Fig. \ref{fig:gaussfit}.}
 \label{tab:correlation_NMF_gaussfit}
 \begin{tabular}{|l|rrrrrrrrrr|}
 \hline
 {} & B1 & B2 & B3 & B4 & B5 & B6 & B7 & B8 & B9 & B10 \\
 \hline
 D1 & 0.21 & 0.51 & 0.03 & 0.08 & 0.09 & 0.05 & 0.38 & {\bf 0.61} & 0.21 & 0.10 \\
 D2 & 0.10 & 0.26 & 0.12 & 0.12 & {\bf 0.34} & 0.18 & 0.14 & 0.15 & 0.17 & 0.25 \\
 D3 & 0.08 & 0.07 & {\bf 0.59} & 0.31 & 0.38 & 0.45 & 0.06 & 0.05 & 0.28 & 0.40 \\
 \hline
 \end{tabular}
 \end{center}
\end{table}

\begin{table}[tb]
 \begin{center}
 \caption{PTEs of spectra B1--B10 in Fig. \ref{fig:XAFS_NMF1} (C). The PTE values were obtained by fitting a Gaussian function to the peaks in spectra B1-B10.}
 \label{tab:PTE_NMF}
 \begin{tabular}{|lr|lr|lr|}
 \hline
 & PTE $[\mathrm{eV}]$ & & PTE $[\mathrm{eV}]$ & & PTE $[\mathrm{eV}]$ \\
 \hline
 B1 & $7730.17$ & B5 & $7729.79$ & B9 & $7729.84$ \\
 B2 & $7730.18$ & B6 & $7729.71$ & B10 & $7729.75$ \\
 B3 & $7729.65$ & B7 & $7730.24$ & & \\
 B4 & $7729.69$ & B8 & $7730.38$ & & \\
 \hline
 \end{tabular}
 \end{center}
\end{table}

Nakamura et al. obtained two-dimensional maps of LC of Li$_x$CoO$_2$ in the model composite electrode from the PTE of the X-ray absorption spectrum at each pixel \cite{RN1483}. Here, we confirm the consistency of the LC map and the domain patterns obtained by the one-sided ONMF. Because the LC can be related to the PTE by a one-to-one function, instead of comparing the LC map and the domain patterns obtained by the one-sided ONMF, we decided to compare the PTE map with the domain patterns.

Figure \ref{fig:gaussfit} (A) shows a two-dimensional map of the PTE obtained from the data on the electrode model charged to 4.2 V at a current rate of 12 mAcm$^{-2}$ in Fig. \ref{fig:gaussfit_full}(A). As described in Section 2.6, three sets of pixels, D1, D2 and D3, into which the pixels were classified with respect to the PTE by thresholds $E_1$ and $E_2$, are plotted in Fig. \ref{fig:gaussfit}(B). B8, B5, and B3 of Fig. \ref{fig:gaussfit}(B) are the domains of the one-sided ONMF that are most highly correlated with D1, D2 and D3. Here, the thresholds maximizing the correlations between the domain patterns of the two methods are $E_1 = 7730.18[\mathrm{eV}]$ and $E_2 = 7729.79[\mathrm{eV}]$. Table \ref{tab:correlation_NMF_gaussfit} is the cross correlation matrix of B1--B10 in Fig. \ref{fig:XAFS_NMF1}(B) and D1--D3 in Fig. \ref{fig:gaussfit}(B). The correlation coefficients between B8 and D1, between B5 and D2, and between B3 and D3 were $0.61$, $0.34$ and $0.59$, respectively. 

Furthermore, to confirm the consistency between domains obtained by the one-sided ONMF and by thresholding the PTE map, we compared the PTEs of the three domains B8, B5 and B3 with the values of thresholds $E_1$ and $E_2$. The PTEs of the spectra of B1--B10 in Fig. \ref{fig:XAFS_NMF1} are listed in Table \ref{tab:PTE_NMF}. The PTE of domain B8, $7730.38[\mathrm{eV}]$, was higher than $E_1 = 7730.18[\mathrm{eV}]$, while the PTE of domain B3, $7729.65[\mathrm{eV}]$, was lower than $E_2 = 7729.79[\mathrm{eV}]$. Moreover, the PTE of domain B5, $7729.79[\mathrm{eV}]$, was equal to $E_2$. Thus, the PTEs of the domains extracted by the one-sided ONMF were correlated with the thresholds $E_1$ and $E_2$ for separating the PTE map into the lower, middle and higher sides. Note that the thresholds $E_1$ and $E_2$ exist in the range where the LC of Li$_x$CoO$_2$ and the PTE satisfy the one-to-one relation.

\subsection{Spectral decomposition of extracted XAS}\label{Sec03-05}
\begin{figure}[htbp]
 \centering
 \includegraphics[scale=.4]{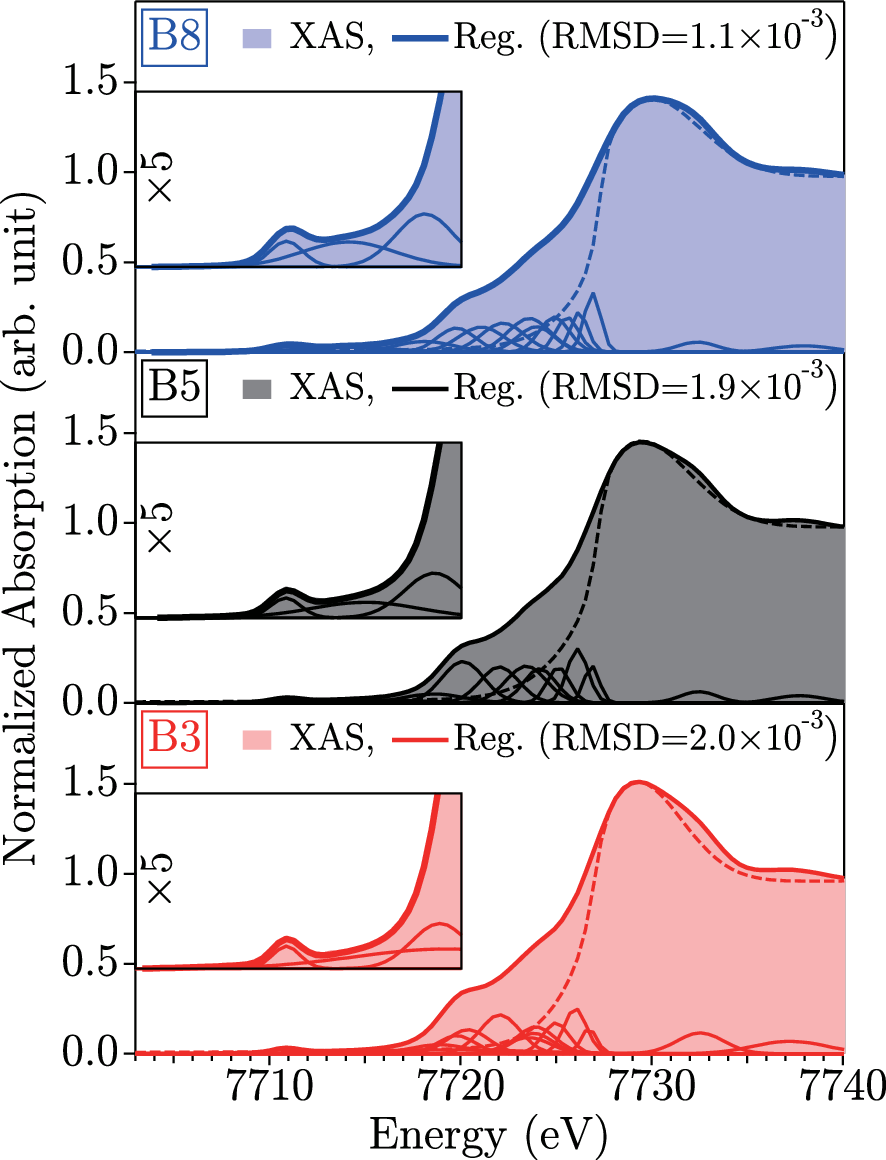}%
 \caption{Spectral decomposition of the extracted XAS for spatial domains B3, B5 and B8. Colored area and thick-solid curve spectra denote the extracted XAS and regressed spectra by $\mathcal{F}_{\hat{K}}(E;\boldsymbol{\theta})$, respectively. Dashed and thin solid curves are $\mathcal{S}(E;\boldsymbol{\theta}_\mathcal{S})$ and $\mathcal{G}_k(E;\boldsymbol{\theta}_k)$ in Eq.~(\ref{iakaiEq01}), respectively. }
 \label{iakaiFig01}
\end{figure}

Figure~7 presents three spatial domain patterns, B3, B5, and B8, which correspond to the spatial patterns D1-3 classified by PTE. We used Bayesian spectroscopy to decompose the XAS extracted from B3, B5, and B8; the results are presented in Fig.~\ref{iakaiFig01}. Colored-area spectra denote the extracted XAS by the one-sided ONMF. Bayesian spectroscopy uses the MAP values, which maximize the total posterior probability. The spectral component of $\mathcal{S}(E;\boldsymbol{\theta}_\mathcal{S})$ and the peak structures of $\mathcal{G}_k(E;\boldsymbol{\theta}_k)$ in Eq.~(\ref{iakaiEq01}) are depicted as dashed and thin solid curves, respectively. By using Bayes free energy minimization, the peaks [$K$ in Eq.~(\ref{iakaiEq01})] are estimated to number 14, 12, and 14 for the XAS extracted from domains B3, B5, and B8, respectively.

The regressed spectra $\mathcal{F}_{\hat{K}}(E;\boldsymbol{\theta})$, which are the sums of these spectral components, are presented by thick solid curves, in which the baseline $y_{0}$ was estimated to be small enough. The root-mean-square deviations (RMSDs) of the $\mathcal{F}_{\hat{K}}(E;\boldsymbol{\theta})$ are in the range of $1.1\textrm{-}2.0\times10^{-3}$ on the normalized absorption intensity scale, and the regressive spectra explained the target XAS well because of the sufficiently small RMSD.

Such fine spectral decomposition provides important information for understanding the charging process and its
spatial inhomogeneity in lithium-ion batteries, because it is considered that the transition energy, intensity,
spectral width, and their changes of each decomposed spectral component reflect the physical properties of the
electronic states involved in ionic conduction.  Indeed, as seen in the insets of Fig.~\ref{iakaiFig01}, one
can find spectral changes in the decomposed components for pre-edge region on the lower energy side of the WL,
in which the ordinate is displayed on a five-times enlarged scale. Here, there are differences in the decomposed
spectral components $\mathcal{G}_k(E;\boldsymbol{\theta}_k)$ ($k=1\sim3$), which will be discussed in
Section~\ref{Sec04-03}.

\section{Discussion}
\subsection{Summary and conclusion}
The automatic domain extraction method that we developed can identify and separate very small differences in X-ray absorption spectra. We removed the reference signal from 2D-XAS data of the model composite electrode by subtracting the X-ray absorption spectrum of Li$_{0.5}$CoO$_2$ reference material from the 2D-XAS data; then we applied the one-sided ONMF to the difference spectra without the reference signal.

We assessed the performance of our proposed method on synthetic data that mimic profiles of X-ray absorption spectra. Bi-cross validation was used to show that number of the ground-truth domains was correctly identified and the spatial domains obtained by our method were identical with the ground-truth domains in the synthetic data (see Figs. \ref{fig:demo_image} and \ref{fig:demo_NMF}). Next, we assessed the performance of our method on two different real 2D-XAS datasets (see Figs. \ref{fig:gaussfit_full}(A) and (B)). We confirmed that our method could determine the rank of the factor matrices $r$ and extract spatial domains with different spectral profiles from these data (see Figs. \ref{fig:XAFS_NMF1} and \ref{fig:XAFS_NMF2}). Then, we quantitatively compared the results obtained with our method and those with the k-means method on the real 2D-XAS data shown in Figs. \ref{fig:gaussfit_full}(A). We confirmed that almost all of the domain patterns obtained with our method were highly correlated with the results obtained with the k-means method (see Figs. \ref{fig:XAFS_NMF1}(B) and \ref{fig:XAFS_KMeans_W} and Table \ref{tab:correlation_NMF_KMeans}); thus, the results of the two methods were consistent. Moreover, we quantitatively compared the results obtained by our method and those by thresholding of the PTE map shown in Figs. \ref{fig:gaussfit_full}(A). We found that the spatial patterns and the PTEs of the domains extracted by the one-sided ONMF were correlated with those of the domains obtained by thresholding the PTE map (see Fig. \ref{fig:gaussfit} and Tables \ref{tab:correlation_NMF_gaussfit} and \ref{tab:PTE_NMF}). Note that the consistency with the results of the k-means method and the PTE map had also been confirmed using other data clipped from a non-active region in Fig. \ref{fig:gaussfit_full}(A) \cite{RN1498}. Finally, we applied Bayesian spectroscopy to spectra of different domains obtained by our method and decomposed the spectra of the individual domains into spectral components. We found that the spectral features of some of the decomposed spectral components differ in the individual domains (see Fig. \ref{iakaiFig01}).

These results lead us to conclude that the one-sided ONMF in combination with removal of the reference signal can extract spatial domains with different spectral profiles from 2D-XAS data showing very little variation from one position to another. Furthermore, we conclude that the individual domains and their spectra obtained by our method represent differences in physically interpretable features, including LC. The physical interpretation of the decomposed spectral components will be discussed below.

\subsection{Advantage over k-means method}
As shown in Fig. \ref{fig:XAFS_KMeans_W} and Table \ref{tab:correlation_NMF_KMeans}, we demonstrated that the k-means method can also automatically extract spatial domains with different profiles of absorption spectra from the 2D-XAS data of the model composite electrode. Our method has the following advantages over the k-means method:

1. The classification achieved by the k-means method is sensitive to the magnitude of each sample, because samples are partitioned into several clusters in which each sample belongs to the cluster with the nearest cluster center. On the other hand, the classification achieved by the NMF is invariant to the magnitude of each sample, because through the matrix factorization the magnitude of each sample is incorporated into the factor matrix $W$ and samples are separated into several clusters in which each sample belongs to the cluster with the most directionally similar column vector in the factor matrix $H$. Thus, our method can achieve magnitude-invariant classification for the spectra.

2. The domain patterns obtained with the k-means method with one-hot labels taking either $1$ or $0$ are not allowed to overlap others, whereas the domain patterns obtained with our method are gently allowed to overlap others because of the column orthogonality imposed on $\boldsymbol{W}$ by using the Lagrange multiplier method (Eq. (8)). Thus, our method can extract spatially overlapped domains.

\subsection{Detection of electronic state changes in different spatial domains}\label{Sec04-03}

\begin{figure}[htbp]
 \centering
 \includegraphics[scale=.4]{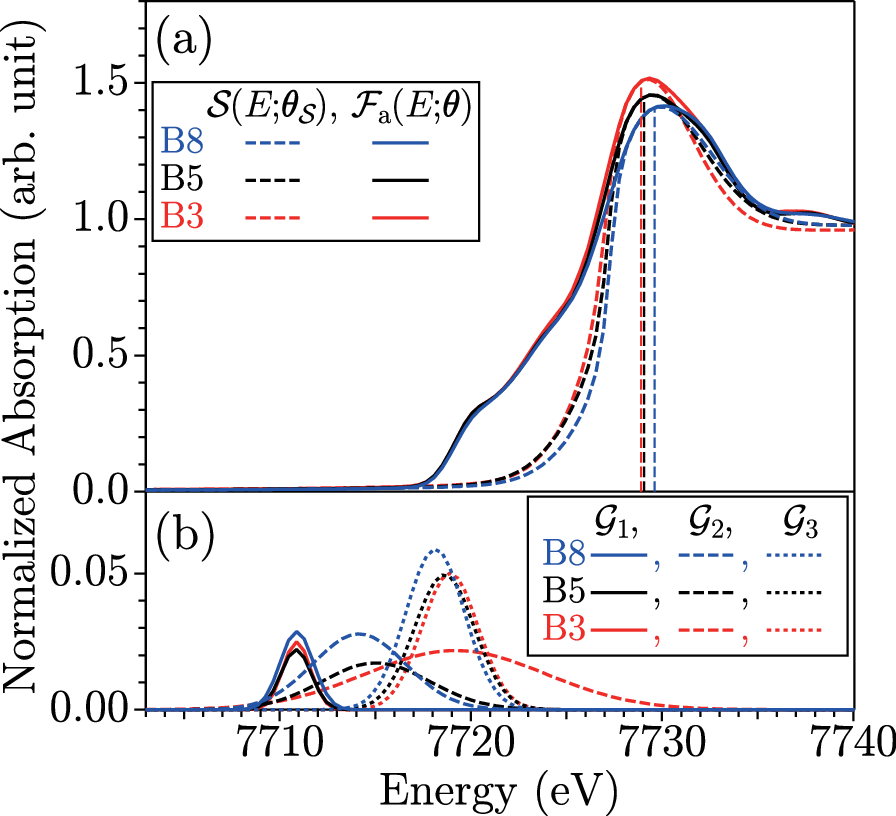}%
 \caption{ (a) Dashed and solid curves mean the spectral components $\mathcal{S}(E;\boldsymbol{\theta}_\mathcal{S})$ and the sum spectra of the high-energy side components defined by Eq.~(\ref{iakaiEq03}), respectively. (b) Spectral components of the pre-edge structures. }
 \label{iakaiFig02}
\end{figure}

As described in Section~\ref{Sec03-05}, the XAS extracted from the spatial domains of B3, B5, and B8 were
decomposed by Bayesian spectroscopy, and differences were found in some of the spectral components. In order to
understand the physical meaning of these differences, the decomposed spectral components are summarized again in
Figs.~\ref{iakaiFig02}(a) and (b).

In Fig.~\ref{fig:gaussfit} of Section~\ref{Sec03-04}, we have shown the domain patterns B8, B5, and B3
corresponding to the two-dimensional maps D1$-$3 with different PTEs.  This change in PTE can be confirmed as a
change in the transition energy of WL, which gives the maximum absorption intensity in XAS through the Bayesian
spectroscopic analysis described in section~\ref{Sec03-05}.
The dashed curves in Fig.~\ref{iakaiFig02}(a) show the spectral components $\mathcal{S}(E;\boldsymbol{\theta}_\mathcal{S})$ including a gentle step and an associated WL peak, as defined by Eq.~(\ref{iakaiEq02}): one can see that their PTEs are different. The vertical dashed lines in Fig.~\ref{iakaiFig02}(a) are the MAP values of the WL-transition energy $E_\textrm{WL}$. The posterior probability distributions of $E_\textrm{WL}$ can be evaluated by using Bayesian spectroscopy on each of the B3, B5, and B8 spatial domains. Their distributions are sharp, and the overlaps among them are not so obvious. Consequently, the change in $E_\textrm{WL}$ between the different spatial domains can be confirmed statistically, and this result supports the result in Section 3.4.

The solid curves in Fig.~\ref{iakaiFig02}(a) are the sum spectra of the high-energy-side components defined by Eq.~(\ref{iakaiEq03}), which include $\mathcal{S}(E;\boldsymbol{\theta}_\mathcal{S})$ and peak structures ($4\leq{}k\leq\hat{K}$) other than pre-edge components ($k=1\sim3$).

\begin{equation}
 \mathcal{F}_\textrm{a}(E;\boldsymbol{\theta}_\textrm{a})
 :=
 \mathcal{S}(E;\boldsymbol{\theta}_\mathcal{S})
 +
 \sum_{k=4}^{K}
 \mathcal{G}_k(E;\boldsymbol{\theta}_k)
 ,
 \label{iakaiEq03}
\end{equation}
where the indexes of $\mathcal{G}_k(E;\boldsymbol{\theta}_k)$ are counted from the low-energy side. Although a noticeable difference appears in the WL, these curves, $\mathcal{F}_\textrm{a}(E;\boldsymbol{\theta}_\textrm{a})$, appear to have no noticeable differences in the low energy region of $E_\textrm{WL}$.

However, we can find spatial inhomogeneities in the electronic states considered to be relating to the ionic
conduction from careful examination of the spectral components of the pre-edge structure decomposed by Bayesian
spectroscopy.  Figure~\ref{iakaiFig02}(b) shows the decomposed spectral components for the pre-edge structures,
which appear on the low energy side of $\mathcal{F}_\textrm{a}(E;\boldsymbol{\theta}_\textrm{a})$.
In Li$_{x}$CoO$_{2}$, Co atoms are located in the crystal field of the octahedral coordination of oxygen atoms~\cite{JPhysCondMat2007V19p436202}, and the peaks depicted by the solid curves, which appear at the lowest energy in Fig.~\ref{iakaiFig02}(b), are considered to be a pre-edge structure due to the transition into the Co-3d $\textrm{t}_\textrm{2g}$ state~\cite{JPhysCondMat2007V19p436202,PRB1992V46p3729,APL2014V104p114105}. However, this transition does not show a significant change in transition energy with respect to the spatial domains B3, B5 and B8.
On the other hand, one can find pronounced changes in the two components decomposed at the high energy side of
Co-3d $\textrm{t}_\textrm{2g}$ as shown by the dashed and dotted curves in Fig.~\ref{iakaiFig02}(b).
It has been reported that the $\textrm{e}_\textrm{g}$ state changes depending on the state of charge~\cite{JPhysCondMat2007V19p436202}. However, it has also been reported that there is a hybridization between the Co-3d $\textrm{e}_\textrm{g}$ state and the 2p orbital of oxygen~\cite{APL2014V104p114105}, and such a difference in hybridization is thought to appear as a change in these two spectral components. The above implies that the electronic-state changes were detected in different spatial domains through the one-sided ONMF of the 2D-XAS data measured on the model composite electrode of Li ion batteries and spectrum decomposed by Bayesian spectroscopy. The method presented here will help us to gain a better understanding of the physical factors governing electrochemical reaction inhomogeneity.

\ack
This work is supported by JST CREST (No. JPMJCR1861).

\newpage 

\bibliographystyle{iopart-num}
\bibliography{./NMF_cite_2}

\end{document}